\documentclass[12pt]{article}
\usepackage{amsmath,amsthm,amsfonts,amssymb}

\def\isom{\cong}

\def\dim{{\hbox{dim}}}

\def\Hom{{\hbox{Hom}}}

\def\a{\alpha}

\def\A{{\cal{A}}}
\def\B{{\cal{B}}}
\def\U{{\cal{U}}}
\def\G{{\mathsf{G}}}

\def\Vir{{\mathrm{Vir}}}
\def\Diff{{\mathrm{Diff}}}
\def\SVir{{\mathrm{SVir}}}
\def\C{{\mathbb{C}}}
\def\R{{\mathbb{R}}}
\def\Z{{\mathbb{Z}}}

\newtheorem{theorem}{Theorem}[section]

\newtheorem{problem}[theorem]{Problem}

\def\SVir{{\mathrm {SVir}}}
\def\C{{\mathbb C}}
\def\R{{\mathbb R}}
\def\Z{{\mathbb Z}}

\title{{\bf From Operator Algebras \\ to Superconformal Field Theory}}

\author{
{\sc Yasuyuki Kawahigashi}\footnote{Supported in part by the
Grants-in-Aid for Scientific Research, JSPS.}\\
Department of Mathematical Sciences\\
University of Tokyo, Komaba, Tokyo, 153-8914, Japan\\
E-mail: {\tt yasuyuki@ms.u-tokyo.ac.jp}}

\begin{document}
\maketitle

\begin{abstract}
We survey operator algebraic approach to (super)conformal
field theory.  We discuss representation theory, classification
results, full and boundary conformal field theories, relations
to supervertex operator algebras and Moonshine,
connections to subfactor
theory of Jones and certain aspects of noncommutative geometry
of Connes.
\end{abstract}

\section{Introduction}
\label{Intro}

Quantum field theory is a physical theory and its mathematical
aspects have been connected to many branches of contemporary
mathematics.  Particularly (super)conformal field theory has
attracted much attention over the last 25 years since \cite{BPZ}.
In this paper we make a review on the current status of
the operator algebraic approach to (super)conformal field
theory.

We start with general ideas of the operator algebraic approach
to quantum field theory, which is called {\sl algebraic
quantum field theory}.  We then specialize conformal field
theories on the 2-dimensional Minkowski space, its light
rays and their compactifications.  We deal with mathematical
axiomatization of physical ideas, and study the
mathematical objects starting from the axioms.  Our
main tool for studying them is representation theory,
and we present their basics, due to Doplicher-Haag-Roberts.
We further introduce the induction machinery, 
called {\sl $\alpha$-induction},
and present classification results.

We then present another mathematical axiomatization of
the same physical ideas.  It is a theory of (super)vertex
operator algebras.  After giving basics, we study the
{\sl Moonshine} phenomena from an operator algebraic viewpoint.

The theory of subfactors initiated by Jones has opened many
surprising connections of operator algebras to various
branches of mathematics and physics such as 3-dimensional
topology, quantum groups and statistical mechanics.
We present aspects of subfactor theory in connection to
conformal field theory.

Noncommutative geometry initiated by Connes is a new 
approach to geometry
based on operator algebras and it is also related to many
branches in mathematics and physics from number theory to
the standard model.  We present new aspects of noncommutative
geometry in the operator algebraic approach to superconformal
field theory.

We emphasize the operator algebraic aspects of (super)conformal
field theory.  Still, we apologize any omission in our discussions 
caused by our bias and ignorance.

\section{Chiral superconformal field theory}
\label{ccft}

\subsection{An idea of algebraic quantum field theory}

We deal with a chiral superconformal field theory, a kind of
quantum field theory, based on
operator algebraic methods in this section.
We start with a very naive idea on what a quantum
field is.  A classical field is simply a function on a
spacetime, mathematically speaking.  In quantum mechanics,
numbers are replaced by operators.  So instead of ordinary
functions, we study operator-valued functions on a spacetime.
It turns out that we also have to deal with something like
a $\delta$-function, we thus deal with operator-valued
distributions, and they should be mathematical formulations
of quantum fields.  An ordinary Schwartz distribution
assigns a number to each test function, by definition.
An operator-valued distribution should assign an operator,
which may be unbounded, to each test function.  There
is a precise mathematical definition of this notion of
an operator-valued distribution, and we can further axiomatize
a physical idea of what a quantum field should be.  Such
an axiomatization is known as a set of {\sl Wightman axioms}.

We work on a Minkowski space ${\mathbb{R}}^4$.
For $x=(x_0,x_1,x_2,x_3)$ and
$y=(y_0,y_1,y_2,y_3)$, their Minkowski inner product is
$x_0 y_0-x_1y_1-x_2y_2-x_3y_3$.
When a linear transform $\Lambda$ on ${\mathbb{R}}^4$
preserves the Minkowski inner product, it is called a
Lorentz transform.  The set of such a $\Lambda$ satisfying
$\Lambda_{00}>0$ and ${\mathrm{det}}\;\Lambda=1$
is called the restricted  Lorentz group.  The universal
cover of the restricted  Lorentz group is naturally
identified with $SL(2,{\mathbb{C}})$.  The set of transformations 
of the form $x\mapsto \Lambda x+a$, where
$\Lambda$ is an element of the restricted  Lorentz group,
is called the restricted Poincar\'e group.
The universal cover of the restricted Poincar\'e group
is naturally written as
$$\{(A, a)\mid A\in SL(2,{\mathbb{C}}),
a\in {\mathbb{R}}^4\}.$$

We say that two regions $O_1, O_2$ are
spacelike separated, if for any
$x=(x_0,x_1,x_2,x_3)\in O_1$ and
$y=(y_0,y_1,y_2,y_3)\in O_2$, we have
$(x_0 -y_0)^2-(x_1-y_1)^2-(x_2-y_2)-(x_3-y_3)^2<0$.

We now briefly explain the Wightman axioms, but we do not go
into full details here, since they are not our main objects
here.  See \cite{SW} for  details.

\begin{enumerate}
\item We have closed operators
$\phi_1(f),\phi_2(f),\dots,\phi_n(f)$ on a Hilbert space $H$
for each test function $f$ on ${\mathbb{R}}^4$
in the Schwartz class.
\item We have a dense subspace $D$ of $H$ which is
contained in the domains of all $\phi_i(f)$, $\phi_i(f)^*$,
and we have $\phi_i(f)D\subset D$ and
$\phi_i(f)^*D\subset D$ for all $i=1,2,\dots,n$.
For $\Phi,\Psi\in D$, the map
$f\mapsto (\phi_i(f)\Phi,\Psi)$ is a Schwartz distribution.
\item We require that
there exists a unitary representation $U$ of
the universal cover of the restricted Poincar\'e group
satisfying the following, where $S$ is an
$n$-dimensional representation of $SL(2,{\mathbb{C}})$ and
$\Lambda(A)$ is the image of $A\in SL(2,{\mathbb{C}})$
in the restricted Lorentz group.
\begin{eqnarray*}
U(A,a)D&=&D,\\
U(A,a)\phi_i(f)U(A,a)^*&=&\sum_j S(A^{-1})_{ij}
\phi_j(f_{(A,a)}),\\
f_{(A,a)}&=&f(\Lambda(A)^{-1}(x-a)).
\end{eqnarray*}
Here the second line means that the both hand sides are equal
on $D$.
\item If the supports of $C^\infty$-functions $f$ and $g$
are compact and spacelike separated, then we have
$[\phi_i(f),\phi_j(g)]_\pm=0$, where the symbol
$[\;,\;]_-$ and $[\;,\;]_+$ denote the commutator and the anticommutator,
respectively, and it means that the left hand side is zero
on $D$.  The choice of $\pm$ depends on $i,j$.
We also have $[\phi_i(f),\phi_j(g)^*]_\pm=0$
\item We have a distinguished unit vector $\Omega\in D$ called the
{\sl vacuum vector}, unique up to phase,
and it satisfies the following.
We have $U(A,a)\Omega=\Omega$, and the spectrum of the
four-parameter unitary group $U(I,a)$, $a\in{\mathbb{R}}^4$,
is in the closed positive cone $$\{(p_0,p_1,p_2,p_3)\mid
p_0\ge 0, p_0^2-p_1^2-p_2^2-p_3^2\ge0\}.$$
We can apply arbitrary $\phi_i(f)$ and $\phi_j(g)^*$ to
$\Omega$ for finitely many times.  We require that
the linear span of such vectors is dense in $H$.
\end{enumerate}

This is a nice formulation and many people have worked
within this framework, but distributions and unbounded
operators cause various technical difficulties in
a rigorous treatment.  Bounded linear operators are much
more convenient for algebraic handling, and we seek for
a mathematical framework using only bounded linear operators.
There has been such a framework pursued by Araki, Haag and
Kastler in early days, and such an approach is called
{\sl algebraic quantum field theory} today.  The basic
reference is Haag's book \cite{Hg} and we now explain its
basic idea as follows.

Suppose we have a family of operator-valued
distributions $\{\phi\}$ subject to the Wightman axioms
as above.
Take an operator-valued distribution $\phi$ and a
test function $f$ on a certain spacetime.
That is, $f$ is a $C^\infty$-function
with a compact support and suppose that its support is
contained in a bounded region $O$ in the spacetime.
Then $\phi(f)$ gives an (unbounded) operator
and suppose it is self-adjoint.
In quantum mechanics, {\sl observables} are represented as
(possibly unbounded) self-adjoint operators.
We now regard $\phi(f)$ as an
observable in the spacetime region $O$.  We have
a family of such quantum fields and many test functions
with supports contained in $O$, so we have many
unbounded operators for each $O$.  Then we make
spectral projections of these unbounded operators
and consider a von Neumann algebra $\A(O)$
generated by these projections.  In this way,
we have a family of von Neumann algebras
$\{\A(O)\}$ on the same Hilbert space
parameterized by bounded spacetime regions.
We now consider what kind of properties such a family
is expected to satisfy, and then we forget the
Wightman fields and just consider these properties
as {\sl axioms} for a family of von Neumann
algebras $\{\A(O)\}$.  Then we mathematically
study such a family
of von Neumann algebras satisfying the axioms.
That is, we construct examples, study relations
among various properties, classify such families, and
so on, just as in a usual axiomatic mathematical theory.

We now list such ``expected properties'' which become
axioms later.  Our spacetime is again
a Minkowski space ${\mathbb{R}}^4$,
where the speed of light is set to be $1$.
Then as a bounded  region $O$, it is enough to consider only
double cones, which are of the form
$(x + V_+)\cap(y + V_-)$, where $x,y\in{\mathbb{R}}^{4}$ and
$$V_\pm = \{ z=(z_0,z_1,\dots,z_3)\in {\mathbb{R}}^{4} \mid
z_0^2 - z_1^2 - z_2^2 - z_3^2 > 0,  \pm z_0 > 0\}.$$

\begin{enumerate}
\item (Isotony) For a larger double cone
$O_2$ than $O_1$, we have more
test functions, and hence more operators.  So we should
have $\A(O_1)\subset \A(O_2)$.
\item (Locality) Suppose two double cones $O_1$ and $O_2$
are spacelike separated.
Then we cannot make any interaction between the two regions
even with the speed of light.  Then observables in the two
regions mutually commute.  We thus require that the
elements in $\A(O_1)$ and $\A(O_2)$ commute.
(This corresponds to the case of the commutator
in the Wightman axioms.)
\item (Poincar\'e Covariance)
In the Minkowski space, the natural symmetry group is the
restricted Poincar\'e group.  We require that
there exists a unitary representation $U$ of
the universal cover of the restricted Poincar\'e group satisfying
$\A(gO)=U_g \A(O) U_g^*$, where $gO$ is the image of $O$ under
the action of the quotient image of $g$ in the Poincar\'e group.
\item (Vacuum) We have a distinguished
unit vector $\Omega\in H$, unique up to
phase, satisfying $U_g \Omega =\Omega$ for all elements $g$ in the
restricted Poincar\'e group,
\item (Cyclicity of the vacuum) We require that
$\bigcup_O \A(O)\Omega$ is dense in $H$.
\item (Spectrum Condition) If we restrict the representation
$U$ to the translation subgroup, its spectrum is contained in
the closure of $V_+$.
\end{enumerate}

It is clear that the above set of axioms is very similar to that
of Wightman.  The assignment of $\A(O)$ to $O$ is traditionally
called a {\sl net} (of von Neumann algebras).

It is very hard to construct an example satisfying the above
axioms.  In the 4-dimensional Minkowski space, we have
only a family of examples known under the
name of {\sl free field models}.

\subsection{Local conformal nets}

Now we work on a little bit different version of a net of
von Neumann algebras explained in the previous subsection.

In the above framework, we have chosen the Minkowski
space ${\mathbb{R}}^4$ as our spacetime and its symmetry
group has been the restricted Poincar\'e group, but it
is also possible to
replace the spacetime and the symmetry group in this
framework.  We now choose the 2-dimensional Minkowski
space and the conformal symmetry group.  We will
describe more details of
this setting in Subsection \ref{fcft}, but for a moment,
we simply regard it as a net $\{\A(O)\}$
of von Neumann algebras satisfying some axioms,
where $O$ is a double
cone in the 2-dimensional Minkowski space.

Then it is possible to ``restrict'' the theory on
two light rays, $\{(t,x)\mid t=\pm x\}$, where
$t,x$ are the time and space coordinates of the
2-dimensional Minkowski space now.  A double cone
is projected to an interval on the light ray, and in
this way, it is possible to have a net of von Neumann algebras
parameterized by intervals on the real line.  We
have a high symmetry group of conformal
transformations, and it is natural and convenient
to consider the one-point compactification $S^1$
of the real line, where $t\in{\mathbb{R}}$
corresponds to $(-t+i)/(t+i)\in S^1$, where $S^1$
is a unit circle on the complex plane and $\infty$
corresponds to $-1\in S^1$.  We now have a net
of von Neumann algebras parameterized by
intervals contained in $S^1$.  
(Se \cite[Section 2]{KL2} for more details on
this ``restriction'' procedure.  See \cite{FRS} and
reference there for this in other approaches to conformal
field theory.)  We now write down
a precise set of axioms.

An interval $I\subset S^1$ means a non-empty, non-dense open
connected subset of $S^1$.  That is, $S^1$ itself nor
$S^1$ minus one point is not an interval.
We have a family $\{\A(I)\}$ of von Neumann algebras on a fixed
Hilbert space $H$ parameterized by intervals $I\subset S^1$.

Note that $PSL(2,{\mathbb{R}})$ acts on $S^1$ through
fractional linear transformation on ${\mathbb{R}}$.
This group is also called the {\sl M\"obius group}.

\begin{enumerate}
\item (Isotony)
For intervals $I_1\subset I_2$, we have $\A(I_1)\subset \A(I_2)$.
\item (Locality) For intervals $I_1, I_2$ with
$I_1\cap I_2=\varnothing$, we have
$[\A(I_1), \A(I_2)]=0$
\item (M\"obius covariance)
There exists a unitary representation $U$ of $PSL(2, \R)$ on $H$
satisfying $U(g)\A(I) U(g)^*= \A(gI)$ for any $g\in PSL(2,\R)$
and any interval $I$.
\item (Positivity of energy)
The generator of the one-parameter rotation subgroup of $U$,
called the {\sl conformal Hamiltonian}, is positive.
\item (Conformal covariance) There exists a projective unitary
representation $U$ of $\Diff(S^1)$ on $H$ extending the unitary
representation of $PSL(2,\R)$ such that for all intervals $I$,
we have
\begin{eqnarray*}
U(g)\A(I) U(g)^* &=& \A(gI),\quad  g\in\Diff(S^1), \\
U(g)AU(g)^* &=&A,\quad A\in\A(I),\ g\in\Diff(I'),
\end{eqnarray*}
where $\Diff(S^1)$ is the group of orientation-preserving
diffeomorphisms of $S^1$ and $\Diff(I')$ is the group of
diffeomorphisms $g$ of $S^1$ with $g(t)=t$ for all $t\in I$.
(Here $I'$ is the interior of the complement of an interval $I$.)
\item (Vacuum vector) There exists a unit
$U$-invariant vector $\Omega$ in $H$,
called the {\sl vacuum vector}, which is cyclic for the
von Neumann algebra generated by $\bigcup_I A(I)$.
\item (Irreducibility) The von Neumann algebra
$\bigvee_{I\subset S^1}\A(I)$ generated by all $\A(I)$'s
is $B(H)$.
\end{enumerate}

It is clear that the above axioms basically correspond to
the axioms for the Minkowski space ${\mathbb{R}}^4$ and
the Poincar\'e group.  Note that the locality now takes
a very simple form.

Now the set of intervals on $S^1$ is not
directed with respect to inclusions, so the name
{\sl net} is not mathematically appropriate, and
the correct terminology should be a {\sl cosheaf},
for example,  but the name ``net'' has been widely used
and we also use it here.  The net $\{\A(I)\}$ satisfying
the above axioms is called a {\sl local conformal net}.

We now list some consequences of the above axioms.
(See \cite{KL1} and references there for more details.)

The vacuum vector $\Omega$ is cyclic and separating
for each von Neumann algebras $\A(I)$.  This is
called the {\sl Reeh-Schlieder theorem}.

Let $I_1\subset S^1$ be the upper semicircle.
Let $C:S^1\to{\mathbb{R}}\cup\{\infty\}$ be
the Cayley transform given by $C(z)=-i(z-1)(z+1)^{-1}$.
We define a one-parameter group $\Lambda_{I_1}(s)$ of
diffeomorphisms by $C\Lambda_{I_1}(s)C^{-1}x=e^s x$.
We also set $r_{I_1}$ by $r_{I_1}(z)=\bar z$ for $z \in S^1$.
For a general interval $I$, we choose $g\in PSL(2,{\mathbb{R}})$
with $I=gI_1$.  Then we set $\Lambda_I=g\Lambda_{I_1}g^{-1}$
and $r_I=gr_{I_1}g^{-1}$.  These are independent of the choice
of $g$, thus well-defined.  The action of $r_{I_1}$ on
$PSL(2,{\mathbb{R}})$ gives a semi-direct product
$PSL(2,{\mathbb{R}})\rtimes{\mathbb{Z}}_2$.
Let $\Delta_I$ and $J_I$ be the modular operator and
the modular conjugation with respect to $(\A(I),\Omega)$.
We now have an extension of the representation
$U$ appearing in the M\"obius covariance property,
still denoted by $U$, such that $U(g)$ is unitary
[resp.~anti-unitary] when $g$ is orientation preserving
[resp.~reversing].  This $U$ satisfies
$U(\Lambda_I(2\pi t))=\Delta_I^{it}$ and $U(r_I)=J_I$.
This is called the
{\sl Bisognano-Wichmann property} \cite{BGL,FG}.

We the have $\A(I')=\A(I)'$ and this is called the
{\sl Haag duality}.  This follows directly from the above
Bisognano-Wichmann property and it means that locality holds
maximally.

It also follows that
each von Neumann algebra $\A(I)$ is automatically a
factor, and it is of type III$_1$, except for the
trivial case $\A(I)={\mathbb{C}}$.  Actually, for
all known cases, each $\A(I)$ is the unique Araki-Woods
factor of type III$_1$.  So each algebra $\A(I)$ does not
contain any information on conformal field theory.  It is
relative relations of factors $\A(I)$ that contain
information on conformal field theory.

If we have a family $\{I_i\}$ of intervals with
$I\subset\bigcup_i I_i$, then the von Neumann algebra
$A(I)$ is contained in the von Neumann algebra
generated by $\bigcup_i A(I_i)$.  This property
is called {\sl additivity} \cite{FJ}.

We have  a notion of a
stronger version of this additivity as follows.
This holds only for some examples, and does not hold in
general.  Let $I$ be an interval and $x\in I$.  Then
$I\smallsetminus \{x\}$ is a union of two intervals
$I_1$ and $I_2$.  The {\sl strong additivity} means
that $A(I)$ is generated by $A(I_1)$ and $A(I_2)$.
This has some similarity to amenability of a single
operator algebra, and gives a kind of amenability type
condition for a family of von Neumann algebras.

We have another related property called the {\sl
split property} \cite{DL}.
Let $I_1, I_2$ be two intervals in $S^1$
with $\bar I_1 \cap \bar I_2=\varnothing$.  Then we
require that $x\otimes y\mapsto xy$ extends to an
isomorphism from $\A(I_1)\otimes \A(I_2)$ to the von
Neumann algebra generated by $\A(I_1)$ and $\A(I_2)$.
(Note that these the two von Neumann algebras act on
$H\otimes H$ and $H$, respectively.)
It is known that if $e^{-tL_0}$ is of trace class
for all $t>0$, where $L_0$ is the conformal Hamiltonian,
then the split property holds \cite{DLR}.  This trace class
condition is easy to verify for concrete examples.
This holds for all known examples, but may not hold
in general.

It is not easy to construct an example of a local
conformal net (except for the trivial one with
$\A(I)={\mathbb{C}}$ for all $I$).  A family of examples
has been constructed by Buchholz-Mack-Todorov \cite{BMT}
from the $U(1)$-currents.

Another important family has been constructed by A. Wassermann
\cite{W}, and the outline of the construction is as follows.
Let $L(SU(N))$ be the loop group, that is, the set of
$C^\infty$-functions on $S^1$ with values in $SU(N)$, with
pointwise multiplication.  For an interval $I\subset S^1$,
we set $L_I(SU(N))$ to be the set of
$C^\infty$-functions $f$ on $S^1$ with values in $SU(N)$
such that $f(z)={\mathrm{Id}}\in SU(N)$ for $z\notin I$.
Then for each positive integer $k$, we have a vacuum
positive energy representation $\pi$ of $L(SU(N))$, which
is a projective unitary representation on some Hilbert
space $H$ having the vacuum vector $\Omega$.
Then we set $\A(I)$ to be the von Neumann algebra
generated by $\pi(L_I(SU(N)))$.  A similar construction
has been studied for other Lie groups by various people
\cite{TL}.  Projective unitary representation of
${\mathrm{Diff}}(S^1)$ has been constructed by \cite{GW,TL}.

There is also a general construction of a local
conformal net from an even lattice.  We will mention this
in Section \ref{moon}.

We finally list some methods to construct new examples
from known examples (with some property).

\begin{enumerate}
\item If we have two local conformal nets $\{\A(I)\}$ on $H$
and $\{\B(I)\}$ on $K$, we have another local conformal
net $\{\A(I)\otimes\B(I)\}$ on $H\otimes K$.  This
is the tensor product construction.  It is easy to see 
that this gives a local conformal net.
\item For a local conformal net $\{\A(I)\}$ with a nice
representation theory, we can make a crossed product
construction by a finite abelian group $G$.  This
is called a {\sl simple current extension} of $\{\A(I)\}$.
(See \cite[Part II]{BE} for a concrete example, though
it is not called a simple current extension there.)
\item If a group $G$ is contained in the automorphism
group of  a local conformal net $\{\A(I)\}$, we can
make a fixed point net $\{\A(I)^G\}$.  (The notion
of an automorphism will be explained in Section \ref{moon}
in detail.)  This is well-studied for finite groups $G$,
and in such a case, this construction is called the
{\sl orbifold construction}. (See \cite{X4} for details.)
\item If we have two local conformal nets
$\{\A(I)\}$ and $\{\B(I)\}$ with $\A(I)\subset \B(I)$,
we can construct a new local conformal net
$\{\A(I)'\cap \B(I)\}$.  This is called
the {\sl coset construction}.  (See \cite{X3} for details.)
\end{enumerate}

In the constructions 2, 3 and 4 in the above, we actually
have to specify the Hilbert space appropriately.

\subsection{Representation theory}

The fundamental tool to study local conformal nets is their
representation theory.

We take a local conformal net $\A$.
A {\sl representation} $\pi$ of $\A$ is a family $\{\pi_I\}$
of representations of $\A(I)$ on the same Hilbert space
$H_\pi$ with the condition that $\pi_{I_1}$ extends
$\pi_{I_2}$ if $I_1\supset I_2$.

We say that a representation $\pi$ on $H_\pi$
is {\sl diffeomorphism covariant}
if there exists a projective unitary representation $U_\pi$
of the universal cover of $\Diff(S^1)$ on $H_{\pi}$ such that
$$\pi_{g I}(U(g)xU(g)^*) = U_\pi(g)\pi_I(x)U_\pi(g)^*$$
for all $x\in\A(I)$ and
all $g$ in the universal cover, where $gI$ is the image
of $I$ by the natural image of $g$ in $\Diff(S^1)$.
A {\sl M\"obius covariant} representation is also defined
in a similar way.  We sometimes use the terminology
{\sl DHR representation}, which means a diffeomorphism covariant
or M\"obius covariant representation depending on the context,
where DHR stands for Doplicher-Haag-Roberts \cite{DHR}.
The identity map gives a DHR representation of $\A$ on the
initial Hilbert space $H$.  This is called the
{\sl vacuum representation}.  We also have a natural
notion of irreducibility for DHR representations.

We now introduce a numerical invariant of the net $\A$ called
the {\sl central charge}.

The {\sl Virasoro algebra} is the infinite dimensional
Lie algebra generated
by elements $\{L_n \mid n\in\Z\}$ and a central element $c$ with
relations
\begin{equation}\label{vir-alg}
[L_m,L_n]=(m-n) L_{m+n} + \frac{c}{12}(m^3-m)\delta_{m,-n}
\end{equation}
and $[L_n,c]=0$. It is the (complexification of) the unique,
non-trivial one-dimensional central extension of the Lie algebra of
$\Diff(S^1)$.

For this infinite dimensional Lie algebra, we have
notions of a {\sl unitary
representation} which means that we have
$L_n^*=L_{-n}$ in the representation space and
{\sl positive energy} which means the image of $L_0$
is positive.
A projective unitary representation of $\Diff(S^1)$ gives
a unitary representation of the Virasoro algebra.
In any irreducible unitary representation the Virasoro algebra,
the element $c$ is mapped to a scalar and its value is
called the {\it central charge}.  We denote this value by
the same symbol $c$, and it is known that all the possible
values of the central charge is the set
$$\{1-6/m(m+1)\mid m=2,3,4,\dots\}\cup[1,\infty)$$
by \cite{FQS} and \cite{GKO}.
The unitary representation of the Virasoro algebra
arising from the projective unitary representation of $\Diff(S^1)$
decomposes into irreducible representations, all with the same
central charge $c>0$.  We define the central charge of the
local conformal net $\A$ to be this value.

The irreducible unitary representations of the Virasoro algebra
produces Wightman fields on $S^1$ and they provide local
conformal nets $\Vir_c$ for all possible values of $c$
\cite{BS-M}.  (See \cite{KL1} for more details.)

We say that a representation $\pi$ is {\sl localized} in a
interval $I_0$ if we have
$H_\pi=H$ and $\pi_{I'_0}={\mathrm{id}}$. For a given
interval $I_0$ and a representation $\pi$ on a separable Hilbert
space, there is a representation $\tilde\pi$ unitarily equivalent to $\pi$
and localized in $I_0$.  This is because all representations
of a type III factor $\A(I'_0)$ on Hilbert spaces are unitarily
equivalent.
If $\pi$ is a representation localized in $I_0$, then by Haag duality
implies that
$\pi_I$ is an endomorphism of $\A(I)$ if $I\supset I_0$.
The endomorphism $\pi$ is called a
DHR endomorphism localized in $I_0$ \cite{DHR}. The
{\sl (Jones) index} of a representation $\pi$ is
the Jones index $[\A(I):\pi_I(\A(I))]$ of $\pi_I$, if $I\supset I_0$.
(This number is independent of $I$.  See Section \ref{subfactor}
for more on the Jones index.)
The {\it statistical dimension} $d(\pi)$ of $\pi$ is the square
root of this index.
The unitary equivalence $[\pi]$ class of a representation $\pi$ of $\A$
is called a {\sl (superselection) sector} of $\A$.

We now introduce a notion of a {\sl tensor product} of
two DHR representations.  Note that we have an obvious
notion of a tensor product for representations of a group,
but not for DHR representations of a local conformal net.
Two DHR representations give two DHR endomorphisms.  Then they
can be composed and the result is still a DHR endomorphism.
This gives a proper definition of a tensor product of DHR
representations, and with this notion, we have a tensor
category of DHR endomorphisms \cite{DHR}.

For group representations $\pi$ and $\sigma$, the tensor products
$\pi\otimes\sigma$ and $\sigma\otimes\pi$ are obviously
unitarily equivalent, but for DHR endomorphisms $\lambda$ and $\mu$
of $\A$, it is not clear to see the relation between
$\lambda\mu$ and $\mu\lambda$.  It turns out they are
unitarily equivalent, and we have a natural choice of unitary
operator ${\mathrm{Ad}}(\varepsilon(\lambda,\mu))\lambda\mu=\mu\lambda$.
The choices of $\varepsilon(\lambda,\mu)$ give a {\sl braiding}
and the tensor category of the DHR endomorphisms gives a
braided tensor category \cite{FRS}.  (Also see \cite{BE} for
an exposition.)

A finite group has only finitely many irreducible unitary
representations up to unitary equivalence.  In theory of quantum
groups, it sometimes similarly happens that a quantum group
has only finitely many irreducible unitary
representations up to unitary equivalence.   Such finiteness
property is sometimes called {\sl rationality}.
We have an operator algebraic counterpart of this notion
for a local conformal net $\A$ as follows and the property
is called {\sl complete rationality}.

Take a local conformal net $\A$ with split property.
We split the circle $S^1$ into four intervals $I_1, I_2, I_3, I_4$
in the counterclockwise order.  Then we have a subfactor
$(\A(I_1)\vee\A(I_3))\subset (\A(I_2)\vee\A(I_4))'$.
We say that the local conformal net $\A$ is {\sl completely
rational} if the index of this subfactor is finite.
We also call the index value of this subfactor
{\sl $\mu$-index} of the local conformal net $\A$.
(In the original definition in \cite{KLM}, strong additivity
was also assumed, but it has been shown in \cite{LX} that
it follows automatically from the above definition.)

In \cite{KLM}, we have shown that complete rationality
implies that we have only finitely many irreducible
DHR representations up to unitary equivalence and that
all these have finite Jones indices.  Furthermore,
the braiding we have for DHR representations is
nondegenerate \cite{KLM}.  That is, we have a finite
dimensional unitary representation of $SL(2,\Z)$
arising from this braiding \cite{R}.  (The
dimension of this representation is the number of
irreducible DHR representations of a local conformal
net up to unitary equivalence.)  The $\mu$-index of
a local conformal net is
equal to the square sum of the dimensions of the
unitary equivalence classes of the
irreducible DHR representations of the local
conformal net \cite{KLM}.  This number measure the
size of the braided tensor category of the DHR
representations.

Wassermann's examples \cite{W} arising from $SU(N)$ at
level $k$ are completely rational by Xu's computation
of the $\mu$-index \cite{X2}.
The Virasoro nets $\Vir_c$ with $c<1$ are also
completely rational \cite{KL1}.

\subsection{$\alpha$-induction and classification}

If we have two groups $H\subset G$, we can obviously
restrict a representation of $G$ to $H$, and furthermore,
we have a notion of an induced representation which gives
a representation of $G$ from that of $H$.  We have a similar
notion of induction for DHR representations of inclusions
of local conformal nets.  This induction procedure was
first defined in \cite{LR1}, further studied in \cite{X1}
with many interesting examples.  It was named as {\sl
$\alpha$-induction} in \cite{BE}, and further studied
in \cite{BEK1}, \cite{BEK2} in connection
to the methods in \cite{O2}.

Suppose we have an inclusion of local conformal nets
$\A\subset\B$.  (Note that the Hilbert space associated
with $\A$ is a subspace of that associated with $\B$.)
Suppose the Jones index of $\A(I)\subset \B(I)$, which
is independent of $I$, is finite.

Take a DHR endomorphism $\lambda$ of $\A$ which is
localized in a fixed interval $I$.  For a while,
we regard $\lambda$ simply as an endomorphism of
a single factor $\A(I)$.  We now would like to
extend it to a larger factor $\B(I)$ as an
endomorphism.  The inclusion map of $\A\subset \B$
naturally defines a DHR representation of $\A$, so
it gives a DHR endomorphism $\theta$ localized in $I$,
and we again regard $\theta$ as an endomorphism of
$\A(I)$.  The braiding structure of the DHR
endomorphisms produces unitary operators
$\epsilon^\pm(\lambda,\theta)$ with
${\mathrm{Ad}}(\varepsilon^\pm(\lambda,\theta))
\lambda\theta=\theta\lambda$.  (Note that a positive
braiding and a negative one automatically come in 
a pair.)

By a general theory of type III subfactors, we
have an isometry $v\in\B(I)$ with
$vx=\theta(x)v$ for all $x\in\A(I)$ and we also have
$\B(I)=\A(I)v$.  We can then extend an endomorphism
$\lambda$ of $\A(I)$ to $\B(I)$ by setting
$\alpha^\pm_\lambda(v)=\varepsilon^\pm(\lambda,\theta)^* v$,
where $\varepsilon^\pm$ means a positive and a negative
braiding operators.  It is not difficult to see that
this indeed gives an extension of an endomorphism.
The extended endomorphism
$\alpha_\lambda^\pm$ does {\sl not} extend to a
DHR endomorphism of $\B$ in general, and it only gives
a slightly weaker version called a
{\sl soliton endomorphism}, but we do not go into details
on this here.

Suppose that a local conformal net $\A$ is completely
rational in the above setting.
Let $\lambda,\mu$ be representatives of unitary
equivalence classes of irreducible DHR representations,
and regard them as endomorphisms of $\A(I)$ for a
fixed interval $I$.  Now $\alpha_\lambda^\pm$ is an
endomorphism of $\B(I)$, and we set
$$Z_{\lambda\mu}=\dim(\Hom(\alpha_\lambda^+,
\alpha_\mu^-)),$$
where we define
$${\mathrm{Hom}}(\rho_1,\rho_2)=
\{a\in M\mid a\rho_1(x)=\rho_2(x)a\ \mathrm{for}\
\mathrm{all}\ x\in M\}$$
for two endomorphisms $\rho_1,\rho_2$ of a von Neumann
algebra $M$.  In this way, we have a square matrix $Z$ whose
size is the number of unitary equivalence classes of
the irreducible DHR representations of $\A$
and whose entries are nonnegative integers.
We denote the vacuum representation of $\A$ by 0, and
we then have $Z_{00}=1$.
Now one of the main results in \cite{BEK1}, Theorem 5.7 there,
says that the matrix $Z$ commutes with the image of the
unitary representation of $SL(2,\Z)$ arising from
the nondegenerate braiding of $\A$.  Actually,
it is shown in \cite{BEK1}
that this property of $Z$ holds in a more general
situation where we have a (possibly degenerate) braiding
not necessarily arising from a local conformal net.
The graphical methods used for this proof are based on
\cite{O2} and are also used in \cite{BEK2}.

In general, when we have a unitary representation of
$SL(2,\Z)$ arising from a nondegenerate braiding as above,
a matrix $Z$ is called a {\sl modular invariant}
if it satisfies the following three conditions, where
$0$ is the vacuum representation.
\begin{enumerate}
\item $Z_{00}=1$.
\item $Z_{\lambda\mu}\in\{0,1,2,\cdots\}$.
\item The matrix $Z$ commutes with the image of the
unitary representation of $SL(2,\Z)$.
\end{enumerate}

In general, for a given such representation of $SL(2,\Z)$,
we have only finitely many modular invariants $Z$.
For the $SU(N)$ nets with level $k$ constructed by A. Wassermann
\cite{W}, the representations of $SL(2,\Z)$  have been
explicitly known and coincide with the previously known
ones in the context of loop groups or Kac-Moody algebras.
The representations of $SL(2,\Z)$ for the Virasoro nets
$\Vir_c$ with $c<1$ follow from \cite{X3}.  (See \cite{KL1}
for explicit descriptions.)

Summarizing the above, we produce a modular invariant
matrix $Z$ from an inclusion of local conformal nets
$\A\subset \B$ where $\A$ is completely rational and
the Jones index is finite.  (Then $\B$ is automatically
completely rational by \cite{KLM}.)  Suppose only $\A$
is fixed.  Then the above gives a map from the set of
extensions
$\B$ with finite indices to the set of modular
invariant matrices $Z$.
(The finite index property of $\B$ automatically
holds if we have irreducibility condition
$\A(I)'\cap\B(I)=\C$ by \cite{KL1} based on \cite{ILP}.)
This map is not injective nor surjective in general, but
for an explicitly
known unitary representation of $SL(2,\Z)$ arising
from the DHR representations of $\A$, the number of
modular invariants is often small and we can explicitly
write down all of them.  Then this severely restricts
possibility of an extension $\B$.
In the case of $SU(2)$ nets
of level $k$ and the Virasoro nets $\Vir_c$ with $c<1$,
complete lists of modular invariants have been given
in Cappelli-Itzykson-Zuber \cite{CIZ}.  Gannon has
such classification lists for many other cases.
See \cite{Ga} for more details.

If $\B$ is a local conformal net with $c<1$, then it is
automatically an irreducible extension of the Virasoro net
$\Vir_c$.  From the known classification of modular invariants
for $\Vir_c$, we can finally classify all possible extensions
$\B$.  This has been done in \cite{KL1} and the classification
list is as follows.

\begin{enumerate}
\item The Virasoro nets $\Vir_c$ with $c<1$.
\item The simple current extensions of the Virasoro nets with index 2.
\item Four exceptionals at $c=21/22$, $25/26$,
$144/145$, $154/155$.
\end{enumerate}

We refer to \cite{FRS} and
reference therein for the role of modular invariants
in other approaches to conformal field theory.

\subsection{Superconformal nets}

We now extend the above theory to superconformal case.
Recall that in the above set of the Wightman axioms, we have seen
$\pm$ in the (anti) commutator.

Let $\A$ be a local conformal net acting
on a Hilbert space $H$ with the vacuum vector $\Omega$.
A unitary operator $U$ on $H$ is said to be an automorphism
of $\A$ if we have $U\Omega=\Omega$ and
$U \A(I) U^*=\A(I)$ for all intervals $I\subset S^1$.  Such a
unitary $U$ is also called a {\sl gauge unitary}.

A $\Z_2$-grading on $\A$ is $\gamma={\mathrm{Ad}}(U)$, where
$U$ is an involutive gauge unitary.  For such a $\Z_2$-grading
$\gamma$, an element $x\in\A(I)$ for some $I$ is said to
be homogeneous if $\gamma(x)=\pm x$.  For such $x$,
we say the parity $p(x)$ is $0$ [resp. $1$] if
$\gamma(x)=x$ [resp. $\gamma(x)=-x$].
Any element $x$ of $\A(I)$ for some $I$ is uniquely decomposed
as $x = x_0 + x_1$ with $p(x_k)=k$.

A {\sl M\"obius covariant Fermi net} $\A$ on $S^1$ is a
$\mathbb Z_2$-graded net whose the symmetry group
is the covering of the M\"obius group and which satisfes the following
property, which is called {\sl graded locality}.

[Graded locality] For $x \in\A(I_1)$, $y\in\A(I_2)$ with
$I_1\cap I_2=\varnothing$, we have $[x,y]=0$, where $[\;,\;]$ is
a graded commutator.  That is, we have
$[x,y]=xy - (-1)^{p(x)p(y)}yx$
for homogeneous $x,y$.

Note the {\sl Bose subnet} $\A_b$, namely the $\gamma$-fixed point
subnet $\A^\gamma$ of degree zero elements, is local.
If we define a unitary $Z$ by $Z=(1 - iU)/(1 - i)$,
then we have $Z\Omega=\Omega$, $Z^2=U$ and
$\A(I')\subset Z\A(I)'Z^*$.

A {\sl Fermi conformal net} is a M\"obius covariant Fermi net
with an extra property on the representation of the 
symmetry group, which is the covering of $\Diff(S^1)$.
We also have a natural notion of a DHR representation
for a Fermi conformal net.  See \cite{CKL} for details.

We say a Fermi conformal net is {\sl completely rational} when
its Boson part is completely rational.

Now we study the two $N=1$ super Virasoro algebras.
They have even generators $c$, $L_n$, $n\in{\mathbb{Z}}$, and
odd generators $G_r$, $r\in {\mathbb{Z}}+1/2$ or
$r\in {\mathbb{Z}}$, with the following relations.
Here $c$ is the central charge and
the elements $L_n$, $n\in{\mathbb{Z}}$, are the usual
Virasoro generators.
\begin{eqnarray}
\label{N1V}
\lbrack L_m , L_n \rbrack
&=&(m-n)L_{m+n} + \frac{c}{12}(m^3 - m)\delta_{m+n, 0},\\
\lbrack L_m, G_r \rbrack&=&\left(\frac{m}{2} - r\right)G_{m+r},\\
\lbrack G_r, G_s \rbrack&=&
2L_{r+s} + \frac{c}{3}\left(r^2 - \frac{1}{4}\right)\delta_{r+s,0}.
\end{eqnarray}
If $r\in {\mathbb{Z}}+1/2$, then the resulting infinite
dimensional Lie algebra is called the {\sl Neveu-Schwarz algebra},
and if $r\in {\mathbb{Z}}$, then the Lie algebra
is called the {\sl Ramond} algebra.
They together make $N=1$ super Virasoro algebras.

As in the Virasoro algebra case, we have a notion of a
unitary (positive energy) representation.  In an irreducible unitary
representation, the central charge $c$ is mapped to
a scalar, and its value is also called the central
charge.  The set of the possible values of the central
charge is now
$$\left\{\frac{3}{2}\left.\left(1-\frac{8}{m(m+2)}\right)
\right| m=3,4,5,\dots\right\}\cup\left[\frac{3}{2},\infty\right)$$
by \cite{FQS}.
Furthermore, in an irreducible unitary representation,
$\{L_n\}$ and $\{G_r\}$ define operator-valued
distributions as in the Virasoro case, and they produce
a Fermi conformal net $\SVir_c$.  It is called a
{\sl super Virasoro net} with central charge $c$.
A Fermi conformal net
is called a {\sl superconformal net} when it is an
extension of a super Virasoro net $\SVir_c$.

Superconformal nets with $c<3/2$ are classified again with
modular invariants listed by Cappelli \cite{Ca} as follows
\cite{CKL}.
\begin{enumerate}
\item The super Virasoro net with
$c=\displaystyle\frac{3}{2}\left(1-\frac{8}{m(m+2)}\right)$,
labeled with $(A_{m-1}, A_{m+1})$.
\item Index 2 extensions of the above (1), labeled with
$(A_{4m'-1}, D_{2m'+2})$, $(D_{2m'+2}, A_{4m'+3})$.
\item Six exceptionals labeled with
$(A_9, E_6)$, $(E_6, A_{13})$, $(A_{27}, E_8)$, $(E_8, A_{31})$,
$(D_6, E_6)$, $(E_6, D_8)$.
\end{enumerate}

\subsection{Full conformal field theory}
\label{fcft}

We now consider a conformal field theory on the $(1+1)$-dimensional
Minkowski space ${\mathcal{M}}$.

A local M\"obius covariant net $\A$ on ${\mathcal{M}}$ is an
assignment of a von Neumann algebra $\A(O)$ on $H$ to each
double cone $O\subset {\mathcal{M}}$ with the following
properties.

\begin{itemize}
\item (Isotony) For $O_1\subset O_2$, we have
$\A(O_1)\subset \A(O_2)$.
\item (Locality) If $O_1$ and $O_2$ are spacelike separated,
then $\A(O_1)$ and $\A(O_2)$ mutually commute.
\item (M\"obius covariance) There exists a unitary representation
$U$ of the direct product of the
universal cover of $PSL(2,\R)$ and its another copy
on $H$ satisfying $U(g)\A(O)U(g)^* = \A(g O)$
for a double cone $O$ and $g\in\U$, where
$\U$ is a connected neighborhood of the identity of
the direct product of the
universal cover of $PSL(2,\R)$ and its another copy
satisfying $gO\subset{\mathcal{M}}$ for all $g\in\U$.
\item (Vacuum vector) There exists a unit $U$-invariant vector
$\Omega$ which is cyclic for the $\bigcup_{O}\A(O)$.
\item (Positive energy) The one-parameter unitary subgroup of $U$
corresponding to the time translations has a positive generator.
\end{itemize}

Let $\G$ be the quotient of the direct product of the
universal cover of $PSL(2,\R)$ and its another copy
module the relation $(r_{2\pi}, r_{-2\pi})=
({\mathrm{id}}, {\mathrm{id}})$.  Then as in
\cite[Section 2]{KL2},
the net $\A$ extends to a local $\G$-covariant net $\A$ on
$\R\times S^1$.  By extending the above definition,
we can also define a {\sl diffeomorphism covariant}
net on $\R\times S^1$ as in \cite[Section 2]{KL2}.
Such a net is a mathematical realization of a {\sl full
conformal field theory}.

We then have representation theory and classification theory
for such local conformal nets \cite{KL2}.
We again refer to \cite{FRS} and
reference therein for studies of full conformal field theory
in other approaches.

\subsection{Boundary conformal field theory}

We can also formulate boundary conformal field theory in our
framework.  In the above setting of the
$(1+1)$-dimensional Minkowski space $\{(x,t)\mid x, t\in\R\}$,
we now restrict our consideration to the half space
$\{(x,t)\mid t\in\R, x>0\}$.  We consider only double
cones contained in this half space.  Then we can formulate
a local conformal net in this setting as in \cite{LR2}.

We have a classification result for such boundary conformal
field theories for small central charges \cite{KLPR}.

\section{Supervertex operator algebras}
\label{voa}

We have seen the notion of superconformal net, which is
a mathematical axiomatization of a physical idea of
superconformal field theory.  However, a superconformal
net does not give the unique possible axiomatization,
and we have another mathematical axiomatization of the
same physical idea.  It is the notion of
(super)vertex operator algebra and we present it here.

As we mentioned at the beginning, a quantum field should
be some kind of operator-valued distribution on a
spacetime.  In our setting, the spacetime is now
the $1$-dimensional circle $S^1$.  So an operator-valued
distribution has a Fourier series expansion
$\sum_{n\in\Z} a_n z^n$.  In our previous formulation
of a chiral (super)conformal field theory with a
(super)conformal net, we have the conformal Hamiltonian
$L_0$, and the Hilbert space $H$ has an eigenspace
decomposition $H=\bigoplus_{n\ge0} H_n$, where
$H_n$ is the eigenspace for $L_0$ with the eigenvalue
$n$ and the direct sum means an $L^2$-direct sum.
Now we consider only an algebraic direct sum of $H_n$,
and we assume that each vector $u$ in this direct sum
produces an operator-valued distribution
$Y(u,z)=\sum_{n\in\Z} u_n z^{-n-1}$, where $a_n$ is
an operator acting on $H$.  (It is customary to use
the number $n$ in
$u_n$ for the coefficient of $z^{-n-1}$.)  This is
called a {\sl state-field correspondence}, because
a vector, called a {\sl state}, gives an operator-valued
distribution, called a {\sl field}, and
$Y(u,z)$ is called a {\sl vertex operator}.

Based on the above idea, a set of axiomatization of
a {\sl supervertex operator algebra} is given as follows.
(There are several variations of the axioms, but we
take one of the simplest forms, which is parallel to
that of our graded local net.)
(See \cite{FLM}, \cite{K} for full details.)

\begin{enumerate}
\item
The space $V$ is a vector space over $\C$ and it
has a superspace decomposition $V^{(0)}\oplus V^{(1)}$,
where $V^{(0)}$ [resp.~$V^{(1)}$] is an even [resp.~odd]
space.  When $v\in V^{(0)}$ [resp.~$v\in V^{(1)}$],
we write $p(v)=0$ [resp~$p(v)=1$].
\item The map $Y$ is from $V\otimes V\to V((z))$ and 
the image of $u\otimes v$ by $Y$ is written as
$Y(u,z)v=\sum_{n\in\Z}u_n v z^{-n-1}$.  For $u\in V^{(j)}$
and $v\in V^{(k)}$, we have $u_n v\in V^{(j+k)}$.
\item We have a vacuum vector ${\mathbf{1}}\in V$ with
$Y({\mathbf{1}},z)u=u$ and $Y(u,z){\mathbf{1}}|_{z=0}=u$
for all $u\in V$.
\item We have a {\sl conformal element} $\omega\in V$ such that
we have $Y(\omega,z)=\sum_{n\in\Z} L_n z^{-n-2}$ and
$L_n$'s satify the Virasoro relation (\ref{vir-alg})
with some $c\in\C$.
\item The action of $L_0$ on $V$ is diaganonalizable as
$V=\bigoplus_{n\ge0, n\in\Z/2} V_n$, where $V_n$ is the
eigenspace of $L_0$ with eigenvalue $n$ and each $V_n$
is finite dimensional.
\item (Translation) We have $[L_{-1},Y(v,z)]=D_z Y(v,z)$.
\item (Locality)  For $u,v\in V$, there is a sufficiently
large integer $N$ satisfying
$$(z-w)^N Y(u,z) Y(v,w)=(-1)^{p(u)p(v)}(z-w)^N
Y(v,w) Y(u,z).$$
\end{enumerate}

A conformal element is also called a Virasoro element.
We say $V$ is a {\sl vertex operator algebra} when
$V^{(1)}=0$.

When $V/V_{-2}V$ is finite dimensional, we say that
$V$ is {\sl $C_2$-cofinite}.  This has some formal similarity
to complete rationality of a local conformal net.
See \cite{H1}, \cite{H2}, \cite{Z} for more details on this
$C_2$-cofiniteness and its consequences.

We also consider a special element $\tau\in V_{3/2}$
called a {\sl superconformal element}.
Its defining property is that the coefficients $G_r=\tau_{r+1/2}$,
$r\in {\mathbb{Z}}+1/2$, of the
corresponding supervertex operator satisfy the Neveu-Schwarz
relations as in (\ref{N1V}).  We have $\tau_0\tau/2=\omega$.
A supervertex operator algebra with a fixed choice of
a superconformal elements is called an $N=1$ supervertex
operator algebra.

The underlying space $V$ should be a Hilbert space $H$
(before completion) in the framework of superconformal nets,
but in the above set of axioms, we have nothing on the
inner products.  If we have an appropriate
positive definite inner product on $V$,
then we say that the supervertex operator algebra has
{\sl unitarity}.  From a viewpoint of relations to operator
algebras, this is the case we are interesed in.

We also have a notion of a {\sl module} over a vertex
operator algebra.  Basically we consider $v_n w$, where
$v$ is in a vertex operator algebra and $w$ is in another
vector space.  We require certain conditions similar to
the above axioms.

The Kac-Moody algebras and integral lattices are two
basic sources to construct examples.  See \cite{FLM}
and other papers for details.

It is expected that superconformal nets and
$N=1$ supervertex operator algebras are in a bijective
correspondence, at least under some nice additional
assumptions,
since they give different axiomatizations of the same
physical objects, but no such general correspondences
have been known.  Still, if one has some idea, technique
or construction for one of the two, we can often
``translate'' it to the other theory.  We explain some
of them below.

\section{Moonshine and its generalizations}
\label{moon}

We now start with the following very general problem.

\begin{problem}
Suppose a group $G$ is given.  Realize it as the automorphism
group of some algebraic structure in an interesting way.
\end{problem}

This formulation is too vague, needless to say.
One classical concrete formulation of the above is the
inverse Galois problem, which asks for a realization
of a given finite group as the Galois group over ${\mathbb{Q}}$,
and is still open today.

Our main objects of interest here are operator algebras.
So we should take operator algebras as the ``algebraic
structure'' in the above problem, but an infinite dimensional
operator algebra always has a rather large group of inner
automorphisms.  One way to kill such inner automorphisms is
to consider
${\mathrm{Out}}(M)={\mathrm{Aut}}(M)/{\mathrm{Int}}(M)$
for a, say, von Neumann algebra $M$.  Popa and Vaes
\cite{PV} has constructed a II$_1$ factor $M$ with
${\mathrm{Out}}(M)={\mathrm{Aut}}(M)/{\mathrm{Int}}(M)
\isom G$ for any given  finitely presented group $G$.

Another operator algebraic formulation of the above
problem is to ask for a realization of $G$ as the
Galois group for an inclusion $N\subset M$, where
the Galois group means
$${\mathrm{Aut}}(M\mid N)=\{\alpha\in
{\mathrm{Aut}}(M)\mid \alpha(x)=x \mathrm{{\ for
\ all\ }}x\in N\}.$$  Such a realization has been
classically known for any finite group $G$
as follows.  Take a free action of $G$ on the hyperfinite
II$_1$ factor $M$.  (For example, realize $M$ as the
tensor product of $|G|$ copies of a hyperfinite II$_1$
factor and let $G$ act on it as permutations of the
tensor copies.)  Then setting $N=M^G$, the fixed
point subalgebra, realizes $G={\mathrm{Aut}}(M\mid N).$

In this section, we present a different formulation
of the above problem based on operator algebras and
some realization examples, but before doing so, we need
to review a development of the Moonshine conjecture and
its solution in the context of vertex operator algebras.

We have so far discussed some realization of given groups, and
we are here interested in finite groups.  Among the finite
groups, the simple ones are obviously basic objects to study.
Today, classification of finite simple groups is complete, and
the classification list consists of the following groups.
(See \cite{FLM}, \cite{Ga} and references there for details.)
\begin{enumerate}
\item Cyclic groups of prime order.
\item Alternating groups of degree 5 or higher.
\item 16 series of Lie type groups over finite fields.
\item 26 sporadic groups.
\end{enumerate}
The groups in the third category are matrix groups such as
$PSL(n, {\mathbb F}_q)$.  The 26 groups in the last category
are the exceptional objects in the classification, and the
first such example was found by Mathieu in the 19th century.
Among these 26 groups, the largest group in terms of the
order is called the {\sl Monster} group, and its order is
$$2^{46}\cdot3^{20}\cdot5^9\cdot7^6\cdot11^2\cdot13^3\cdot
17\cdot19\cdot23\cdot29\cdot31\cdot41\cdot47\cdot59\cdot71,$$
which is around $8\times10^{53}$.  This group was first
constructed by Griess \cite{G} as the automorphism group of
a certain 196884-dimensional commutative nonassociative algebra.
The smallest dimension of Monster's non-trivial irreducible
representation is known to be 196883.

We now recall the definition and properties of the classical
{\sl $j$-function}.  It is a function of a complex
number $\tau$ with ${\mathrm{Im}}\;\tau>0$ and defined as
\begin{eqnarray*}
j(\tau)&=&\frac{(1+240\sum_{n>0}\sigma_3(n)q^n)^3}
{q\prod_{n>0}(1-q^n)^{24}}\\
&=&q^{-1}+744+196884q+
21493760q^2+864299970 q^3+\cdots,
\end{eqnarray*}
where $\sigma_3(n)$ is the sum of the cubes of the divisors
of a positive integer $n$ and $q=\exp(2\pi i \tau)$.

We have the modular invariance property,
$j(\tau)=j\left(\displaystyle\frac{a\tau+b}{c\tau+d}\right)$ for
$\left(\begin{array}{cc}
a & b \\ c& d\end{array}\right)\in SL(2,{\mathbb{Z}})$, and this
is the only function satisfying this property and starting with $q^{-1}$,
up to freedom of the constant term.  The constant term 744 above is
rather arbitrary, and we here use $J(\tau)=j(\tau)-744$.

McKay noticed $196884=196883+1$ for the first nontrivial coefficient
of the function $J(\tau)$ and the first nontrivial dimension of
the irreducible representations of the Monster group.  This might
look purely accidental, but similar relations for the
other coefficients of the $J$-function and the dimensions of
the irreducible
representations of the Monster group have been subsequently found.
Based on these observations,
Conway-Norton \cite{CN} formulated
the {\sl Moonshine conjecture} roughly
as follows, which has been now proved by Borcherds \cite{B}.

\begin{enumerate}
\item
We have a ``natural'' infinite dimensional graded vector space
$V=\bigoplus_{n=0}^\infty V_n$ with $\dim V_n<\infty$
having some algebraic structure whose automorphism group
is the Monster group.
\item
For each element $g$ in the Monster, the power series
$\sum_{n=0}^\infty ({\mathrm{Tr}}\;g|_{V_n})q^{n-1}$ is a
special function called a {\sl Hauptmodul} for
some discrete subgroup of $SL(2,{\mathbb{R}})$.
When $g$ is the identity element, we obtain
the $J$-function.
\end{enumerate}

The power series in the second part above is called the
{\sl McKay-Thompson series}.  The discrete subgroups
appearing in the second part have a special property
called {\sl genus zero property}.
The statement in the first part is vague, since it does not
specify ``some algebraic structure''.
It was in response to to this problem that
Frenkel-Lepowsy-Meurman \cite{FLM} gave an axiomatization
of vertex operator algebras.  They also gave a realization
of $V$ in the above part (1) and called it
the {\sl Moonshine vertex operator}, which is written as
$V^\natural$, since it should be a {\sl natural} structure.
We sometimes call the property of the vertex operator
algebra in the second part above
the {\sl Moonshine property}.

Their construction roughly goes as follows.
They start with the exceptional lattice in dimension 24 called
the Leech lattice $\Lambda$.  This is a special embedding of
${\mathbb{Z}}^{24}$ into ${\mathbb{R}}^{24}$, and the
inner product of any two vectors in the image is always an
even integer.  (There is a deep theory on such lattices.
See \cite{CS} for example.)  Then they have a general
construction of a vertex operator algebra $V_\Lambda$
for this lattice $\Lambda$.  Physically speaking, it corresponds
to a theory of strings living on ${\mathbb{R}}^{24}/\Lambda$.
The involution sending $x\to -x$ on $\Lambda$ induces an
automorphism of $V_\Lambda$ of order 2.  We take a fixed
point vertex operator algebra of this automorphism, then
it turns out that this has a nontrivial extension, and
this extended vertex operator algebra is the Moonshine
vertex operator algebra $V^\natural$.  (This extension at the
last step is given by the simple current extension.)
This construction
is called a {\sl twisted orbifold construction}, where
the name ``orbifold'' refers to the fixed points of an action
of a finite group.

Miyamoto \cite{M} has a new construction of $V^\natural$
based on the fact that it is an extension of the 48th tensor power
of the Virasoro vertex operator algebra $L(1/2,0)$, which was
found by \cite{DMZ}.  This kind of extension of tensor powers
of the Virasoro vertex operator algebra $L(1/2,0)$ has been
studied in the name of {\sl framed vertex operator algebra}
by \cite{DGH}.

We have constructed an operator algebraic counterpart
$\A^\natural$ of the Moonshine vertex operator algebra
based on this idea of framed vertex operator algebra
in \cite{KL4} as follows.

The Virasoro vertex operator algebra $L(1/2,0)$ has a direct
counterpart ${\mathrm{Vir}}_{c=1/2}$ as a local conformal net.
The representation theory of ${\mathrm{Vir}}_{c=1/2}^{\otimes k}$
is well-understand, so we can make their extensions as
local conformal nets.  In this way, we can construct
$\A^\natural$ naturally.
(Dong-Xu \cite{DX} has a general construction of local
conformal nets from lattices, as a counterpart of general
lattice vertex operator algebras.)

The Hilbert space on which the local conformal net $\A^\natural$
acts is simply a Hilbert space completion of $V^\natural$ with
its natural positive definite inner product, and the Virasoro
algebra has a unitary representation with $c=24$ on it.   The Virasoro
generator $L_0$ is the generator of the rotation group and
it has the eigenspace decomposition $H=\bigoplus_{n\ge0} H_n$
where $H_n$ is the eigenspace of $L_0$ with eigenvalue $n$.

The group of all gauge unitary operators is the automorphism
group of the net $\A$ and it is
also called the {\sl gauge group} of the net $\A$.  It is
always a compact group.
Such a unitary operator $u$ automatically commutes with the
action of the M\"obius group, so in particular, it preserves
the decomposition $H=\bigoplus_{n\ge0} H_n$.
(Note that such a unitary operator automatically acts on the Virasoro
subnet ${\mathrm{Vir}}_c$ trivially if the local
conformal net is strongly additive by \cite{CW}, which is
the case we are mainly interested.)
So the McKay-Thompson series for the local conformal net
$\A^\natural$ are identified with those for the vertex operator
algebra $V^\natural$.

Now we have to prove that the automorphism group of $V^\natural$
and that of $\A^\natural$ are identified.  It is easy to see
that the former is contained in the latter, but the converse
inclusion is nontrivial.  The vertex operator corresponding to
the conformal element is called the {\sl stress-energy tensor}, and
it is naturally interpreted as an operator-valued distribution
on the circle.  In the case of $V^\natural$ as an extension of
$L(1/2,0)^{\otimes48}$, we have 48 copies of such stress-energy
tensors with $c=1/2$ and their automorphic images under the Monster
group action are also nice operator-valued distributions, since
each such automorphism is a unitary operator acting on our
Hilbert space.  In this way, we have ``sufficiently many''
operator-valued distributions, and from this fact, we can prove
that each automorphism of the net $\A^\natural$ indeed arises
from an automorphism of the vertex operator algebra $V^\natural$
as in \cite[Theorem 5.4]{KL4}.

We now discuss other finite simple groups.
Among the 26 sporadic finite simple groups, we have three
groups with Conway's name attached.  They are $Co_1$, $Co_2$,
$Co_3$ and $Co_1$ has the largest order, around
$4.2\times 10^{18}$.
It is isomorphic to the automorphism group of the
Leech lattice divided by its center of order 2.
Duncan constructed an ``super'' analogue of the Moonshine vertex
operator algebra and showed that its automorphism group,
as an $N=1$ supervertex operator algebra, is this group $Co_1$
in \cite{D1}.  We now present its operator
algebraic counterpart.  We study this object within a general
operator algebraic framework for super conformal field theory.

Duncan considered an $N=1$ supervertex
operator algebra in this setting.  An automorphism of
a supervertex
operator algebra fixing the superconformal element
is said to be an automorphism of an $N=1$ supervertex
operator algebra.

Duncan constructed two $N=1$ supervertex operator algebras
$A^{f\natural}$ and $V^{f\natural}$, and showed they are
isomorphic in \cite{D1}.  He then showed
its group of automorphisms of $N=1$ supervertex operator
algebra is Conway's group $Co_1$.  Its character
${\mathrm{Tr}}(q^{L_0-c/24})$ is
\begin{equation}
\label{char-co}
q^{-1/2}+276q-{1/2}+2048q+\cdots=
\frac{\theta_{E_8}{\tau}\eta(\tau)}
{\eta(\tau/2)^8\eta(2\tau)^8}-8,
\end{equation}
where $q=\exp(2\pi i\tau)$, ${\mathrm{Im}}\;\tau>0$.
The construction of $V^{f\natural}$ is a twisted
${\mathbb Z}_2$-orbifold construction from a $N=1$
supervertex operator algebra
arising from the lattice ${\mathbb Z}^4\oplus E_8$ as in
discussions after Theorem 6.1 in \cite{D1}.
It is an extension of $L(1/2,0)^{\otimes24}$, where
$L(1/2,0)$ is the Virasoro vertex operator algebra with
$c=1/2$.  We first have an analogue of \cite[Lemma 5.1]{KL4}.
That is, we first consider a vertex operator subalgebra
of $V^{f\natural}$ generated by $g(L(1/2,0)^{\otimes24})$ for
all $g\in Co_1$.  Then it turns out that this is the
even part of the supervertex operator algebra $V^{f\natural}$.
We next consider the supervertex operator algebra generated
by the even part of the supervertex operator algebra
$V^{f\natural}$ and its superconformal element.  
Since the even part is a fixed point of an automorphism
of order 2, the Galois correspondence shows that
this supervertex operator algebra must be equal to $V^{f\natural}$
itself.

The representation theories of the vertex operator algebra
$L(1/2,0)$ and the local conformal net ${\mathrm{Vir}}_{1/2}$
are identified on the level of Hilbert spaces as in
\cite[Section 3]{KL1} based on \cite{X3}.  Let $H$ be
the Hilbert space completion of $V^{f\natural}$ with respect
to the natural inner product on the extension of
$L(1/2,0)^{\otimes24}$.  Then $k$th
stress-energy tensor $T_k(z)$, $k=1,2,\dots,24$, with $c=1/2$
acts on $H$ as in the arguments after Lemma 5.1 in
\cite{KL4}.  Let $G(z)$ be the superstress-energy tensor with
$c=12$ arising from the superconformal element.  As in
\cite[Lemma 5.2]{KL4},
\cite[Section 6]{CKL}, \cite[Sections 4--5]{CHKL},
the family of Wightman fields
$$\{gT_k(z)\mid g\in Co_1,
k=1,2,\dots,24\}\cup\{G(z) \}$$
are strongly graded local, where the definition of
strong locality in \cite[Section 5]{KL4} is extended
to the strongly graded local case.  Note that each
$g\in Co_1$ gives a unitary operator on $H$ and
$gG(z)g^{-1}=G(z)$ by the definition of the automorphism group
of an $N=1$ supervertex operator algebra.

Now as in \cite[Lemma 5.2]{KL4}, we have a graded local
net $\{\A^{f\natural}(I)\}$ having subnet
$\{{\mathrm{SVir}}_{c=12}(I)\}$.  As in \cite[Theorem 5.4]{KL4},
we can show that the automorphism group of the graded local net
$\{{\mathrm{SVir}}_{c=12}(I)\}$ and the automorphism group
of the supervertex operator algebra $V^{f\natural}$ leaving
the natural inner product invariant are
identified.  From the above construction, the subgroup
of the latter fixing the superconformal element is identified
with the subgroup of the former fixing the subnet
$\{{\mathrm{SVir}}_{c=12}(I)\}$  pointwise.  (Each element
of $Co_1$ fixes the natural inner product of $V^{f\natural}$
by the construction \cite{D1}.)

\begin{theorem}
\label{superM}
The superconformal net $\A^{f\natural}$ constructed above is
a completely rational graded local 
net with $c=12$ having ${\mathrm{SVir}}_{c=12}$
as a subnet.  Its character is given by (\ref{char-co}) and
the group of automorphisms of $\A^{f\natural}$
fixing ${\mathrm{SVir}}_{c=12}$ pointwise is Conway's group
$Co_1$.
\end{theorem}

Note that the McKay-Thompson series for each element
in $Co_1$ has been computed by Duncan in \cite[Section 7]{D1},
and we have the same series in the operator algebraic approach.

A similar structure has been pursued for other sporadic finite
simple groups also by Duncan \cite{D2}.  It is known that 20 of
the 26 sporadic finite simple groups are ``involved''
in the Monster in the sense that they are quotients of
subgroups of the Monster group.  The Conway groups
$Co_1$, $Co_2$, $Co_3$ are among these 20, and the other
six are called ``pariah'' groups.
One of the ``pariah'' groups is the Rudvalis group $Ru$,
and its order is around $1.5\times 10^{11}$.  It is closely
related to the  Conway-Wales lattice of rank 28 over
${\mathbb{Z}}[i]$.

Duncan constructed two supervertex operator algebras with
automorphic actions of the Rudvalis group with certain analogue
of the Moonshine property on two-variable power series arising from
group elements of the Rudvalis group in \cite{D2}.
We now construct an operator algebraic
counterpart for one of the two.  The other supervertex operator
algebra of Duncan in \cite{D2} has no unitarity, so
it has no operator algebraic counterpart.  (That is, we
do not have a positive definite inner product, so we
cannot construct a Hilbert space from the very beginning.)

Duncan \cite{D2} has an ``enhanced supervertex operator
algebra'' $A_{Ru}$.  Now we first ignore the ``enhanced
structure'', then it is simply an $N=1$ supervertex operator
algebra with $c=28$ containing
$L(1/2,0)^{\otimes56}$ as in the above case of $A^{f\natural}$.
The above machinery to construct a graded local net from
56 copies of stress-energy tensors with $c=1/2$ and their
automorpshic images under the action of the Rudvalis group
together with a single super stress-energy tensor with $c=28$
produces a completely rational superconformal net $\A_{Ru}$,
with the Rudvalis group $Ru$ acting as the automorphisms
fixing the $N=1$ super Virasoro subnet
$\{{\mathrm{SVir}}_{c=28}(I)\}$ elementwise.  Let
$\{\B(I)\}$ be the fixed point net of $\{\A_{Ru}(I)\}$
with the action of the Rudvalis group.  Then by the
classical Galois correspondence, the group of
automorphisms of $\{\A_{Ru}(I)\}$ fixing $\{\B(I)\}$
elementwise is the Rudvalis group.  The subnet
$\{\B(I)\}$ should correspond to the mysterious
{\sl enhanced} structure in \cite{D2}, but the meaning
is not understood well yet.

The analogue of the Moonshine property of Duncan \cite{D2}
makes sense together with $\tilde\forall_{Ru}$, but this
supervertex operator algebra does not have unitarity, so
the operator algebraic interpretation of this property is
rather incomplete, unfortunately.

In the above constructions, we have dealt with separate
groups separately.  From traditional ideas in operator
algebras, all amenable objects should have some unified
constructions.  In our setting, the groups are finite,
so they are amenable, needless to say, and von Neumann
algebras are amenable, that is, injective if we have
a split property which is known to hold in all the above
examples.  So we expect some uniform construction which
works for all finite groups at once, and it would give
a much deeper understanding on vertex operator algebras,
but such a construction seems far from today's
understanding, unfortunately.

\section{Subfactor theory}
\label{subfactor}

Now we discuss relations of the above framework
to general theory of subfactors.
Jones initiated his theory of subfactors \cite{J1} first
for type II$_1$ factors.  Kosaki \cite{Ko} extended it to
arbitrary factors, and Longo \cite{L1} showed that the
statistical dimension of a superselection sector in
the Doplicher-Haag-Roberts theory \cite{DHR} is identified
with the square root of the Jones index for the image of
an endomorphism of a type III factor.  This is how subfactor
theory is related to quantum field theory, and a great deal
of interactions have been worked our over many years.
Here we make a quick review on classification theory.

By Popa's deep analytic results in \cite{P}, classification
of subfactors of the hyperfinite II$_1$ factor $M$ with
finite Jones index is reduced to classification of certain
representation theoretic invariants if we have certain
amenability condition on the subfactor.  The case of
{\sl finite depth}, where we have certain finiteness of
irreducible objects in a tensor category of representations,
has caught much attention.  This gives a special case of
amenable subfactors, and roughly similar to the
{\sl rational} case of a conformal field theory and related
to the theory of quantum groups at roots of unity.
Such representation theoretic data are characterized by
various methods such as Ocneanu's paragroup \cite{O1}
and Jones' planar algebras \cite{J2}.  A fundamental invariant
is the principal graph of a subfactor, which is a finite
graph for the case of finite depth.

From the beginning of the subfactor theory \cite{J1},
it has been known that the index value 4 is the first special
value.  Classification of subfactors with index less than 4
was found by Ocneanu \cite{O1} and the case with index equal
to 4 was also worked out by various people.  If the index
is less than 4, the principal graph must be one of the
$A_n$-$D_{2n}$-$E_{6,8}$ Dynkin diagrams. For each of
$A_n$ and $D_{2n}$ graphs, we have a unique subfactor, and
for each of $E_6$ and $E_8$, we have two subfactors.
We refer to \cite{EK} for details.

Haagerup considered a problem of listing subfactors up to
index $3+\sqrt3$ in \cite{Ha}.  Up to this index value, if
a subfactor does not have a finite depth, then the
principal graph must be $A_\infty$.  He gave a list of
countable graphs, and showed that if we have a subfactor
with index value in $(4,3+\sqrt3)$, then the principal
graph must be one in the list.  He and Asaeda gave
realization of two in the list in \cite{AH}.
One infinite series in \cite{Ha} were shown to be impossible
in Bisch \cite{Bi}, and we had no progress on the remaining cases
for some years.

Then Etingof, Nikshych and Ostrik \cite[Theorem 8.51]{ENO} proved that
the Jones index of a subfactor with finite depth must be a cyclotomic
integer.  That is, the index value is an algebraic integer contained
in the field ${\mathrm{Q}}(\zeta)$, where $\zeta$ is some root of unity.
Asaeda and Yasuda \cite{A}, \cite{AY} proved that this
kills all of the remaining graphs in the Haagerup list \cite{Ha}
except for one.  This final remaining case has been realized
recently by Bigelow, Morrison, Peters and Snyder \cite{BMPS}.

The proof of Etingof, Nikshych and Ostrik \cite{ENO} is not
easy to understand for operator algebraists, so here we present a
version of their proof in a style more familiar to operator algebraists.

The starting point is the following result of Coste-Gannon
\cite{CG}, where $N_{ij}^k$ is a nonnegative integer
defined by the Verlinde formula
\begin{equation}
\label{verlinde}
N_{ij}^{k}=\sum_l \frac{S_{il}S_{jl} \overline{S_{kl}}}{S_{0l}}.
\end{equation}

\begin{theorem}[Coste-Gannon]
Let $(S_{ij})_{i,j=0,1,\dots,m}$ be
a symmetric unitary matrix with the following
properties.
\begin{eqnarray*}
&&N_{ij}^k\equiv\sum_l\frac{S_{il}S_{jl}\bar S_{kl}}{S_{0l}}
{\mathrm{\ is\ rational\ for\ all\ }}i,j,k,\\
&&S_{0j}>0, \quad{\mathrm{ for\ all\ }} j.
\end{eqnarray*}
Then we have a cyclotomic field $F$ containing all $S_{ij}$.
\end{theorem}

The proof of the above result actually
shows commutativity of the Galois
group for the Galois extension of ${\mathbb{Q}}$ containing
all $S_{ij}$.  Then the classical Kronecker-Weber theorem
gives that the Galois extension is contained in some cyclotomic
field.  (A proof for this is also included in the appendix
of \cite{ENO}.)

We now start a proof of the statement that the Jones index is a
cyclotomic integer, if we have a finite depth.
Let $N\subset M$ be a subfactor with finite index and finite depth.
We may assume that $N$ and $M$ are of type III, by tensoring a
common type III factor if necessary.  (This is not essential.
The following arguments can be easily translated into the
bimodule language for type II$_1$ subfactors.)  Suppose that
$\{\rho_i\}_{i=1}^n$ is a system of irreducible
endomorphisms of $M$ arising
from the subfactor $N\subset M$.  Then we obtain the Longo-Rehren
subfactor \cite{LR1}, $M\otimes M^{\mathrm{opp}}\subset R$, where we have
$\bigoplus_{i=1}^n \rho_i\otimes\rho_i^{\mathrm{opp}}$ as
the dual canonical endomorphism for this subfactor.  This is
a ``quantum double subfactor'' for the system $\{\rho_i\}_{i=1}^n$
and we follow the description in \cite{I}.  Note
that we have a system $\{\rho_i\otimes\rho_j^{\mathrm{opp}}\}_{i,j}$
of irreducible endomorphisms of $M\otimes M^{\mathrm{opp}}$.
Let the system $\{\lambda_k\}_{k=1}^m$ of irreducible endomorphisms of $R$
be the one arising from $\{\rho_i\otimes\rho_j^{\mathrm{opp}}\}_{i,j}$
and the subfactor $M\otimes M^{\mathrm{opp}}\subset R$
as in \cite[Section 4]{I}.  By \cite[Theorem 5.5]{I}, the
system $\{\lambda_k\}_{k=1}^m$ gives a modular
tensor category and we have the Verlinde formula as above by \cite{R}.
Then by the above theorem of Coste-Gannon, we have a cyclotomic
field $F$ which contains all $S_{ij}$ arising from the system
$\{\lambda_k\}_{k=1}^m$.  In particular, we have
$d(\lambda_k)=S_{0k}/S_{00}\in F$ for all $k$, where $d(\lambda_k)$
stands for the statistical dimension, which is equal to
$[R:\lambda_k(R)]^{1/2}$.
Let $\iota$ be the inclusion map for the subfactor
$M\otimes M^{\mathrm{opp}}\subset R$.  Denote the index value
$[R:M\otimes M^{\mathrm{opp}}]$ by $w$.  Note that
$w$ is a sum of some $d(\lambda_k)$'s with multiplicity, so it is
in the field $F$.  Then for any $j$, we have a decomposition
$[\iota(\rho_j\otimes {\mathrm{id}})\bar\iota]=
\bigoplus_l l_{jk} \lambda_k$, where $l_{jk}$ is the multiplicity.
Then we have
$$d(\rho_j)=d(\rho_j\otimes {\mathrm{id}})=
\frac{\sum_k l_{jk} d(\lambda_k)}{w}\in F.$$
Now the canonical endomorphism $\gamma_M$ for the subfactor
$N\subset M$
decomposes as $\bigoplus_i n_i \rho_i$, where $n_i$ is the
multiplicity.  Then we have
$$[M:N]=\sum_i n_i d(\rho_i)\in F,$$
which gives the desired conclusion.

\section{Noncommutative geometry}

Now we discuss relations to noncommutative geometry of Connes
\cite{C}.

A commutative unital $C^*$-algebra is isomorphic to $C(X)$ where
$X$ is a compact Hausdorff space.  So a general $C^*$-algebra is
regarded as a noncommutative analogue of a compact Hausdorff
space, but in order to study geometry, we need more structure
than just a compact Hausdorff space.

Let $M$ be a closed Riemannian
manifold.  From the $C^*$-algebra $C(M)$,
we can recover $M$ only as a topological space, so even in the
commutative case, we need additional structures.  If the manifold
has an extra structure called a {\sl spin structure}, we
have a spinor bundle on $M$, and its $L^2$-sections give
a Hilbert space $H$.  The smooth function algebra $C^\infty(M)$
acts on this Hilbert by a pointwise multiplication, and
we have an unbounded self-adjoint operator $D$ on this $H$,
called the {\sl Dirac operator}, which is a kind of
a ``square root'' of the Laplacian on $M$.  From the triple
$(C^\infty(M), H, D)$, we can recover complete geometric
information on $M$.  From these, the Connes axiomatization
of a  {\sl noncommutative compact
Riemannian spin manifold} is given as a triple
$({\mathcal A}, H, D)$ of a $*$-subalgebra $\A$ of $B(H)$ for
a Hilbert space and a self-adjoint operator $D$ on $H$.
Such a triple is called a {\sl spectral triple}.

\begin{enumerate}
\item All the resolvents of $D$ are compact operators.
\item We have $[D,a]\in B(H)$ for all $a\in{\mathcal A}$.
\end{enumerate}

The commutator $[D,a]$ has a domain naturally, and we
mean that it has a bounded extension.

If a spectral triple arises from a compact Riemannian
spin manifold as above, then
the condition that the dimension of the manifold
is less than $p$ is
expressed in terms of the eigenvlues of the Laplacian
as the condition that the
$n$th eigenvalue of $L_0$ is $O(n^{-1/p})$.
We have an infinite dimensional version of this condition
called {\sl $\theta$-summability} which is defined
by the condition ${\mathrm{Tr}}(e^{-t D^2})<\infty$
for all $t>0$.

Longo \cite{L3} suggested relations between
superselection sectors in conformal field theory
and elliptic operators.  Conformal field theory should
give an infinite dimensional version of a noncommutative
manifold, because it has infinite degree of freedom.
Also see \cite{FGw} for connections of superconformal
field theory and noncommutative geometry.

For an ordinary Riemannian manifold, the asymptotic
behavior of ${\mathrm{Tr}}(e^{-2\pi t\Delta})$ as $t\to0+$
contains geometric information on the manifold.  In \cite{KL3},
we have pursued a similar study for the asymptotic behavior
of $\log({\mathrm{Tr}}\;e^{-2\pi tL_0})$ for a local conformal
net.  These two asymptotic behaviors analogous, but note that
we have ``$\log$'' for the latter.  This comes from the
``infinite dimensionality'' of our noncommutative structure.
Anyway, here we have some correspondence between the Dirac
operator $\Delta$ of a Riemannian manifold and the conformal
Hamiltonian $L_0$ of a local conformal net.
So we expect that the Dirac operator is somehow analogous
to a square root of $L_0$.
One of the Ramond relations (\ref{N1V}) gives $G_0^2=L_0-c/24$,
and $-c/24$ simply gives a scalar in a representation, so
if we have an $N=1$ supersymmetry, we expect that the image
of $G_0$ gives an analogue of the Dirac operator as a part
of a spectral triple.

Based on this analogy,  nets of spectral triples
$({\mathfrak A}(I), H, D)$
parametrized by intervals $I\subset S^1$ have been
constructed in \cite{CHKL}.  We first have a graded local
net $\{\A(I)\}$ first from a representation of the
Ramond algebra on $H$, but we have to drop the axiom on
the vacuum vector, since representations of the Ramond
algebra do not have a vacuum vector. Then $G_0$ gives the ``Dirac
operator'' on the same Hilbert space $H$.  We now need
a $*$-algebra for a spectral triple.  We have a super
derivation $\delta=[\cdot,D]$, where the bracket means the
supercommutator.  Let ${\mathrm{Dom}}(\delta)$ be the
set of operators $x\in B(H)$ with $\delta(x)\in B(H)$ in
an appropriate sense.  Then we have $C^\infty(\delta)=
\bigcap_{n=1}^\infty {\mathrm{Dom}}(\delta^n)$.
For each interval
$I\subset S^1$, we can set
${\mathfrak A}(I)=\A(I)\cap C^\infty(\delta)$,
and we do obtain a net of spectral triples, but the
problem is that this ${\mathfrak A}(I)$ may be too small,
e.g., it may be that we have ${\mathfrak A}(I)={\mathbb{C}}$.
It has been actually shown in \cite{CHKL} that
${\mathfrak A}(I)$ is strongly dense in $\A(I)$ for
each interval, so each ${\mathfrak A}(I)$ is certainly
nontrivial.

\noindent {\bf Acknowledgements.} The authors thanks 
John F. Duncan  for explanations on his work, and
Sebastiano Carpi and the referee for comments on 
this paper.
\end{document}